\documentclass[12pt]{article}
\usepackage{t1enc}
\usepackage[latin1]{inputenc}
\usepackage [dvips]{graphicx}
\usepackage{flafter}
\usepackage[dvips]{rotating}
\usepackage{latexsym}
\begin{document}
\begin{center}{\bfseries \large  Does telomere elongation lead to a longer lifespan if cancer is considered?}
\end{center}
\centerline{Michael Masa$^1$, Stanis{\l}aw Cebrat$^2$ and Dietrich Stauffer$^1$ 
\footnote{Corresponding author. Email: stauffer@thp.uni-koeln.de \ Fax: +49-221-470-5159}} 
\bigskip
\noindent
$^1$ Institute for Theoretical Physics, Cologne University, D-50923 K\"oln,
Euroland \
\noindent
\newline
$^2$ Institute of Genetics and Microbiology,
University of Wroc{\l}aw,\\ ul. Przybyszewskiego 63/77, PL-54148 Wroc{\l}aw,
Poland \
\begin{abstract}
As cell proliferation is limited due to the loss of telomere repeats in DNA of normal
somatic cells during division, telomere attrition can possibly play an important role in
determining the maximum life span of organisms as well as contribute to the process of 
biological ageing. 
With computer simulations of cell culture development in organisms, which consist of
tissues of normal somatic cells with finite growth, we otain an increase of life span
and life expectancy for longer telomeric DNA in the zygote.
By additionally considering a two-mutation model for carcinogenesis and indefinite proliferation by
the activation of telomerase, we demonstrate that the risk of dying due to cancer can outweigh 
the positive effect of longer telomeres on the longevity.
\end{abstract}
\begin{normalsize}
\end{normalsize}
\section{Introduction} 
Telomeres are tandem repeated noncoding sequences of nucleotides at both ends of the DNA in
eukaryotic chromosomes stabilizing the chromosome ends and preventing them
from end-to-end fusion or degradation \cite{blackburn}. Polymerase cannot completely replicate
the 3' end of linear DNA, so telomeres are shortened at each DNA replication \cite{harley}.
This end replication problem leads to a finite replicative capacity for normal somatic cells
\cite{olovnikov}.
They can only divide up to a certain threshold, the Hayflick limit \cite{hayflick01,hayflick02}.
The enzyme telomerase, repressed in most normal somatic cells, synthesizes and elongates
telomere repeat sequences at the end of DNA strands so that certain cells like germline
cells are immortal and indefinite in growth~\cite{greider,shay01}.

Most forms of cancer follow from the accumulation of somatic mutations \cite{nordling, vogelstein}.
Cancer-derived cell lines and 85-90\% of primary human cancers are able
to synthesize high levels of telomerase and thus are able to prevent further shortening of
their telomeres and proliferate indefinitely~\cite{kim}. 
But if cells are premalignant or already cancerous and telomerase is not yet activated,
the proliferative capacity of these cells and therefore the accumulation of mutations is 
determined by the remaining telomere length~\cite{moolgavkar01}.
So the frequency of malignant cancer should be higher for longer telomeres in normal 
somatic cells.

Recently published data show that longer telomeric DNA increased the life span of nematode worms
\cite{joeng}. Experiments are running~\cite{lanza} whether there is a positive effect on the 
longevity also for organisms with renewing tissues if the telomere length in fetal cells is
increased \cite{lanza}. 
As the probability for the incidence of cancer is correlated to the replicative potential of
the mutated cells \cite{shay02}, one can ask the following question: 
Is an extension of life span possible if telomeres in embryonic cells are elongated and cancer 
is considered?
An answer to this question could be given by computer simulations as the presented model
focuses on organismal ageing due to the loss of telomeres in DNA and neglects other effects which
lead to a decreasing survival probability with age.

As shortening of telomeres is one of the supposed mechanisms of ageing on cellular level,
most stochastical and analytical studies investigate this relationship~\cite{levy01,arino,
olofsson,rubelj,sozou}. A theoretical model which directly relates telomere attrition to human ageing
was first suggested by Aviv et al.~\cite{aviv}.
Here we present a different telomere dynamics model providing requirements to study the effects of
(a) different mean telomere lengths in constituting cells, as well as of (b) somatic mutations 
leading to cancer progression and indefinite proliferation due to telomerase activity on the
ageing of organisms.

\section{Basic model of biological ageing due to telomere attrition}
\begin{figure}[!ht]
\centering
\includegraphics[width=14.0cm]{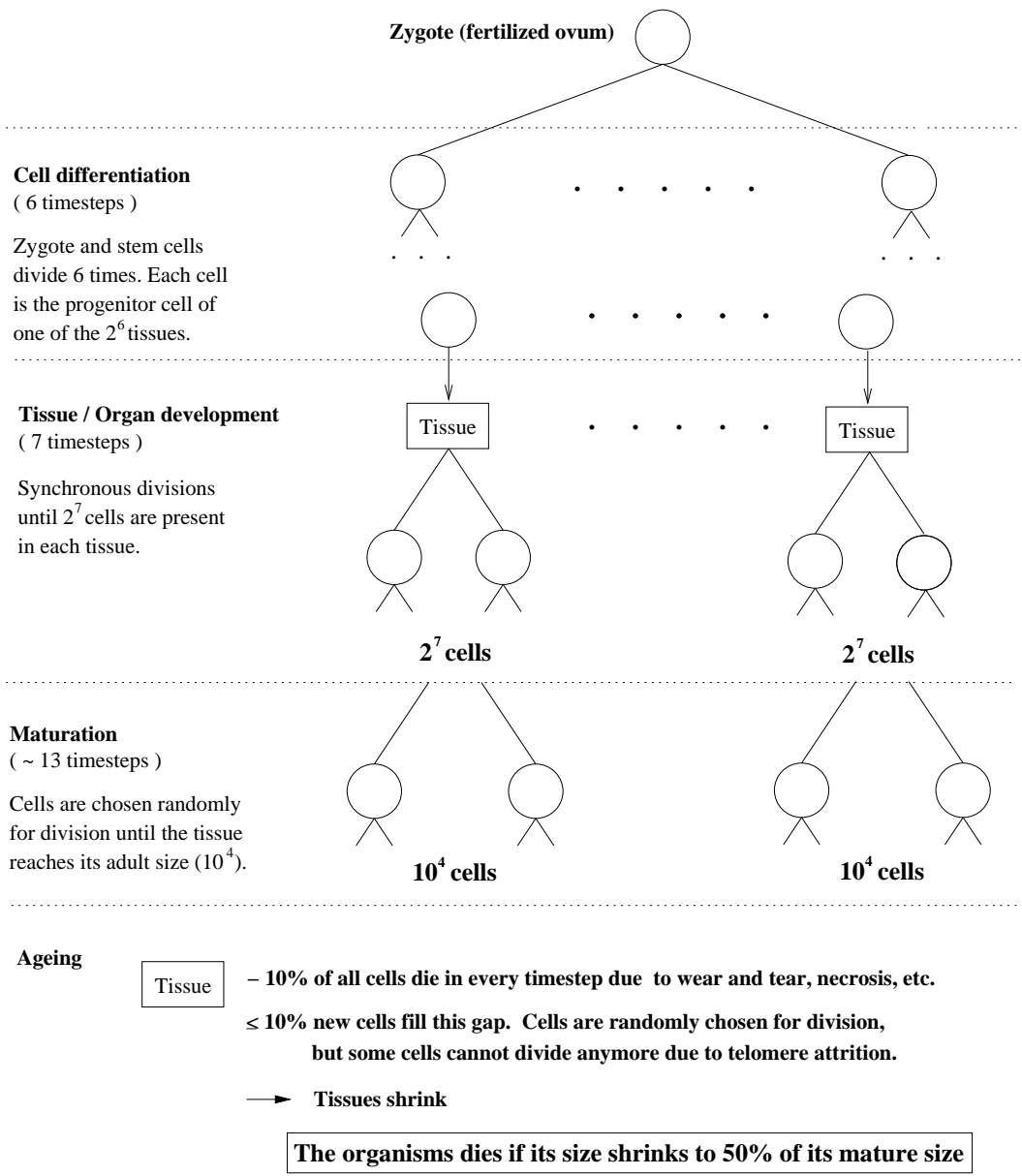}
\caption{Dynamics of cell proliferation in one organism in the basic model}
\label{bild01}
\end{figure}

Every organism is developed from a single progenitor cell, the zygote
(figure~\ref{bild01}).
The initial telomere length of zygote cells is assumed to be normally distributed with mean
$\mu_{z}$ and standard deviation $\sigma_{z}$ \cite{buijs}.
Telomere repeats lost per division (TRLPD) are randomly chosen at each division of every cell
from a normal distribution with mean $\mu_{TRLPD}$ and standard deviation $\sigma_{TRLPD}$.
A dividing cell produces a clone who inherits the replicative capacity of the progenitor
cell at this age.
Cells can divide until nearly all their original telomeres are lost.

For every organism the dynamics of the model is as follows: Divisions of the zygote
and the stem cells derived from it occur 6 times in the early embryo.
Each of these cells is the progenitor of one tissue. This is followed by a period 
of population doublings where all cells divide once in every timestep
until $2^7$ cells are present in each tissue.
In the following maturation stage, cells are chosen randomly for division until each tissue
reaches the adult size of $10^4$ cells. It takes about 26 timesteps until an organism is mature.

Ageing starts now. In every timestep first cells die with 10\% probability due to events like necrosis
or apoptosis. 10\% of the cells of the corresponding tissue are then randomly chosen for division to fill
this gap. The replacement does not have to be complete as the chosen cell could probably not divide
anymore due to telomere attrition. After some time the tissue will start shrinking. 
The random choice of dying and dividing cells in differentiated tissues is in accordance with
nature as for example in epithelium the choice of cells to be exported from the basal layer is
random \cite{cairns,frank}. 
The organism dies, if its total cell population size shrinks to 50\% of the mature size.
The results presented in the following do not depend qualitatively on the choice of this threshold.
\section{Results without cancer}
\begin{figure}[!ht]
\centering
\includegraphics[width=9cm,angle=-90]{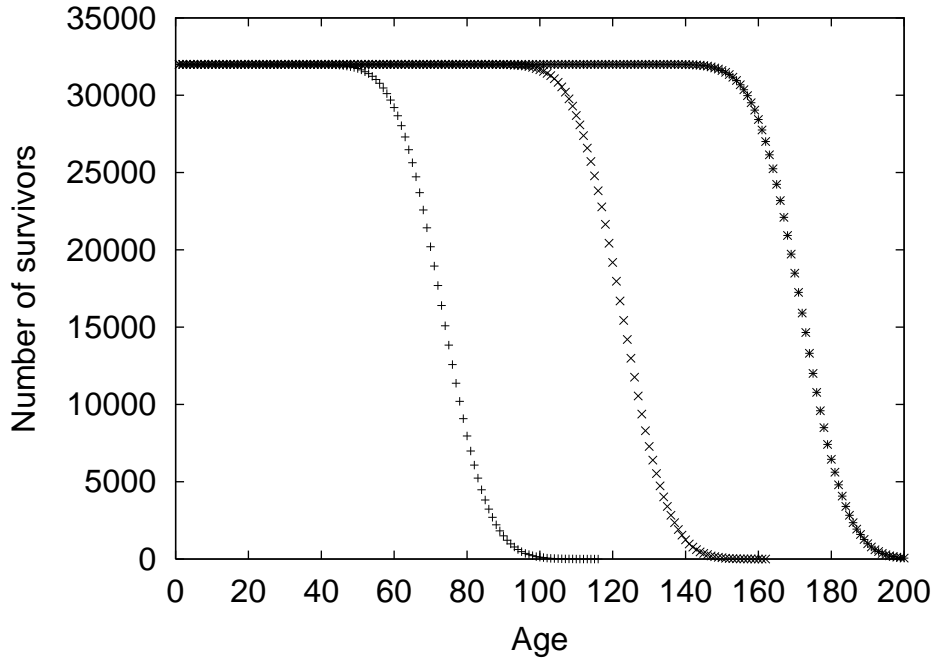}
\caption{Age distribution of 32000 organisms with telomere lengths of $\mu_z=1500(+)$,
 $\mu_z=2000$($\times$) and $\mu_z=2500(*)$; $\sigma_z=100$, $\mu_{TRLPD}=50$, $\sigma_{TRLPD}=10$}
\label{bild02}
\end{figure}

In the implementation of this model, linear congruential generators are used to produce the required
random numbers and normally distributed variables are generated with the Box-Muller algorithm \cite{tu,box}.
Resulting age distributions for different mean telomere lengths in the zygote cells are shown in
(figure \ref{bild02}).
The shape of these distributions is analogous to empirical data of many human and animal
populations. We obtain a positive effect on the longevity of the organisms if the mean
telomere length in the precursor cell is increased.
The chosen mean doubling potentials of the zygote cells are 30, 40 and 50 with the choice
of $\mu_z=1500,2000,2500$ and $\mu_{TRLPD}=50$. The number of mitotic divisions observed
in human fibroblasts is higher \cite{allsopp01}, but the choice of this parameter
is reasonable as the number of considered cells per mature organism (640000) in this
model is also much lower than in human organisms where the total number of cells is of the order
of $10^{13}$ \cite{cairns}. Non-dividing cells are not included in the model as we focus on ageing
due to the progressive shrinking of tissues driven by telomere shortening. Our
results will be given below as part of  Fig.5.

\section{Introducing carcinogenesis and telomerase}
Clonal cancer is now introduced in the model. In accordance to the model of Moolgavkar et al.,
one of our assumptions is that malignant tumors arise from independently mutated progenitor cells
\cite{moolgavkar02}.
For most forms of carcinoma, transformation of a susceptible stem cell into a cancer cell is
suggested to be a multistage process of successive mutations with a relatively low probability
for the sequential stages \cite{nordling,armitage01}.
Two independent and irreversible hereditary mutation stages are considered here, which can occur
at every level of development of the organism during cell division.

The first premalignant stage to be considered is a promotion stage:
A dividing cell can mutate with small probability $p_{mut}$ \cite{cairns}.
All descendant cells inherit this mutation.
This mutation leads to a partial escape from homeostatic control of growth by the local cellular
environment \cite{moolgavkar01,sarasin}.
Cells on the promotion stage have a selective advantage over unaffected cells \cite{armitage02}.
In our model they are chosen first for division during maturation and for filling up the gap in
the ageing period.

The subsequent transition can occur again with probability $p_{mut}$ during division. If a cell
reaches this second stage of mutation it is a progenitor of a carcinoma.
An explosive clonal expansion to a fully malignant compartment happens \cite{sarasin}.
This cell and the clonal progeny doubles in the current timestep until it is no more possible
due to telomere attrition.
This expansion leads only to an increase of the malignant cell population size.
As a certain fraction of cells is killed per unit time and clonal expansions only occur with a
very small probability, the tumor environments may not continue growing, eventually shrink,
or even die \cite{hiyama02}. We assume that it is necessary for advanced cancer progression and
therefore for the development of a deadly tumor that fully mutated cells are able to
activate telomerase \cite{shay03}. 

In our model, telomerase activation is possible at every age of the organism in normal and
mutated cells during division with a very low probability $p_{telo}$. The irreversible loss
of replicative potential is stopped in these cells.
As the contribution of telomerase to tumorigenicity is not yet completely understood~\cite{hiyama,kanjuh,blasco}, we assume that death of an organism due to cancer occurs if
telomerase is reactived in at least one fully mutated cell \cite{moolgavkar03}. 
We treat the time interval between the occurence of the deadly tumor and death as
constant, so we set this interval to zero.
\section{Effects of different telomere lengths considering cancer}
\begin{figure}[!h]
\centering
\includegraphics[width=9cm,angle=-90]{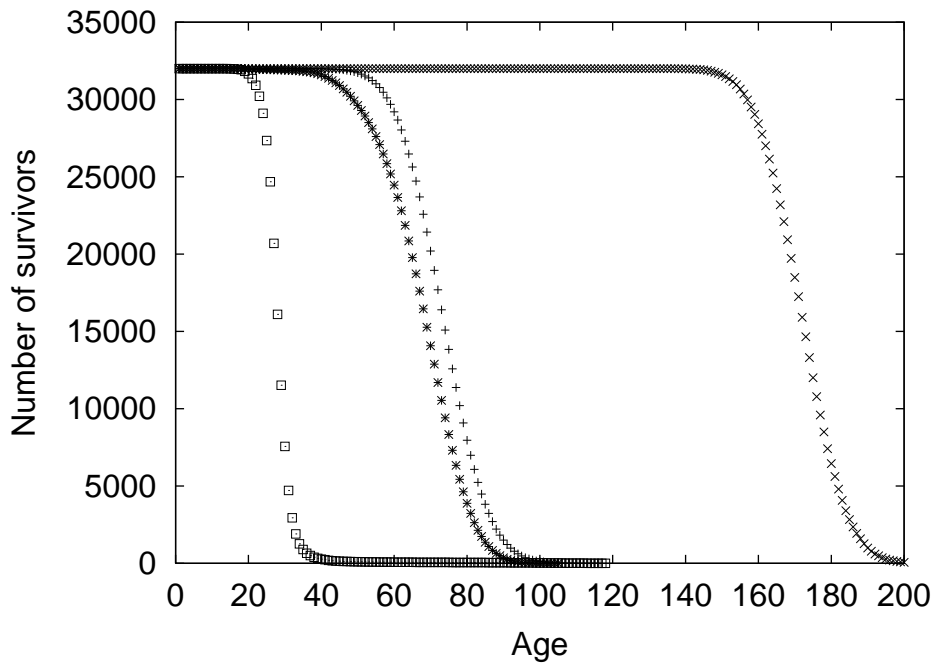}
\caption{Age distribution of 32000 organisms with telomere lengths of $\mu_z=1500$ with ($*$) and without cancer (+) and $\mu_z=2500$ with ($ \Box $) and without cancer ($\times$); $\sigma_z=100$,
$\mu_{TRLPD}=50$, $\sigma_{TRLPD}=10$. Cancer mutations are possible with $p_{mut}=5*10^{-5}$.
 Telomerase can be activated with $p_{telo}=10^{-5}$.}
\label{bild03}
\end{figure}

Figure \ref{bild03} shows simulation results for $\mu_z=1500$ and $\mu_z=2500$. 
 As we consider a lower complexity by chosing a lower number of tissues and cells per
organism, we assume higher mutation rates for the incidence of cancer than observed in 
nature \cite{luebeck,drake}.
The age distribution for shorter initial telomere lengths considering cancer is shifted to
the left but still very old organisms exist. For longer telomeres the age distribution is again
shifted to the left but even behind the distributions for shorter telomeres with and without
considering carcinogenesis.
\begin{figure}[!h]
\centering
\includegraphics[width=9.0cm,angle=-90]{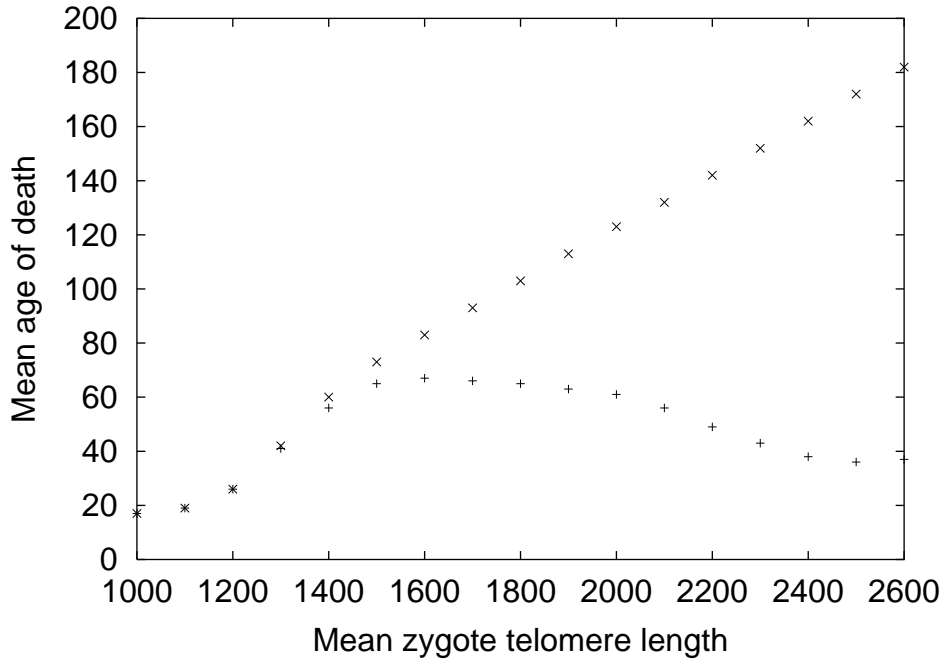}
\caption{Mean age of death for different inital zygote telomere lengths $\mu_z$ (32000 organisms)
         without cancer ($\times$) and introducing carcinogenesis ($+$).} 
\label{bild04}
\end{figure}

Thus without considering cancer, organisms with longer zygote telomeres live longer,
as the life expectancy of the organisms increases linear for longer telomeres.
But if cancer is considered this effect is reversed for longer initial mean telomere lengths 
(fig. \ref{bild04}).
\begin{figure}[!ht]
\centering
\includegraphics[width=9cm,angle=-90]{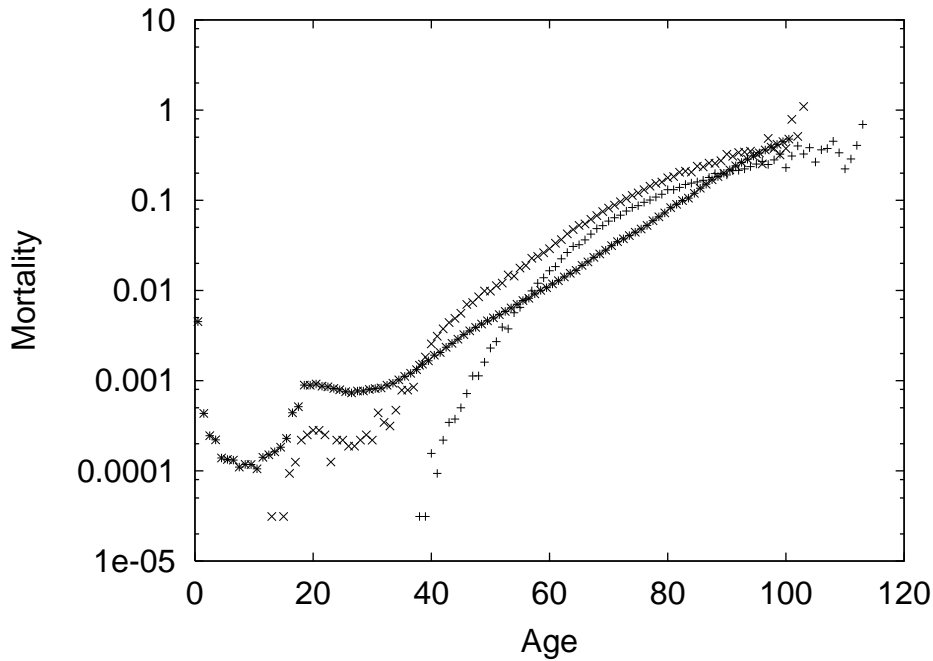}
\caption{Mortality function for $\mu_z=1500$ with ($\times$) and without cancer (+), 32000 organisms considered, $\sigma_z=100$, $\mu_{TRLPD}=50$, $\sigma_{TRLPD}=10$; German men ($*$) from www.destatis.de (June 2004) (Sterbetafel 2000/2002)}
\label{bild05}
\end{figure}

The force of mortality~\cite{thatcher01} resulting from this model is shown for $\mu_z=1500$
with and without considering cancer in comparison to empirical human mortality 
data (figure~\ref{bild05}). 
It agrees to some extent with human mortality functions provided cancer is incorporated into
the model and decelerates at advanced ages, as claimed for human and animal populations 
\cite{vaupel01,vaupel02}. The hump in the curve at younger ages, occurring also for other
parameter sets, fits to data of many human mortality tables.

\section{Conclusion}
The expected simulation result of the basic model without cancer is an increase of life span of
most organisms with longer initial telomeres. After introducing somatic mutations promoting cancer
and telomerase activation in this model, the survival probability is lower for each considered
initial telomere length in certain time intervals in adult ages.

But even low probabilities for the two mutation stages and for the activation of telomerase lead
to a strong reduction of life span for longer telomeres.
So the implication of two-stage carcinogenesis for the incidence of cancer in this simple model
of cell proliferation in organisms is that life expectancy and life span of complex organisms
cannot be increased by artificially elongating telomeres in primary cells, for example during
a cloning procedure.

Further improvements, extentions and applications of this model are possible. With respect to
the role of telomeres and telomerase in carcinogenesis, maybe this computational approach can
contribute to the development of a comprehensive theoretical model in oncology uniting mutagenesis
and cell proliferation \cite{gatenby}.
\\ \\
{\bf Acknowledgements}
\\
We wish to thank the European project COST-P10 for supporting visits of MM and DS 
to the Cebrat group at Wroc{\l}aw University and the Julich supercomputer center
for computing time on their CrayT3E. CS was supported by Foundation for Polish
Science.

\end{document}